\documentstyle[aps,epsf,12pt]{revtex} 
\tightenlines 
\draft 
\narrowtext 
 
\begin{document} 
 
\title 
{Heisenberg model with Dzyaloshinskii-Moriya interaction: \\ 
A mean-field Schwinger boson study} 
 
\vskip 0.7cm 
 
\author{L. O. Manuel, C. J. Gazza, A. E. Trumper, \\ and H. A. Ceccatto} 
 
\address 
{Instituto de F\'{\i}sica Rosario, Universidad Nacional de Rosario\\ 
and Consejo Nacional de Investigaciones Cient\'{\i}ficas y T\'ecnicas,\\ 
Boulevard 27 de Febrero 210 Bis, 2000 Rosario, Rep\'ublica Argentina }

\maketitle

\vskip 1.cm 
 
\begin{abstract}  
We present a Schwinger-boson approach to the Heisenberg model with  
Dzyalo\-shinskii-Moriya interaction. We write 
the anisotropic interactions in terms of Schwinger bosons keeping the 
correct symmetries present in the spin representation, which allows us 
to perform a conserving mean-field approximation. Unlike previous studies of 
this model by linear spin-wave theory, our approach takes into account  
magnon-magnon interactions and includes the effects of  
three-boson terms characteristic of noncollinear phases. The results reproduce  
the linear spin-wave predictions in the semiclassical large-$S$ limit, and  
show a small renormalization in the strong quantum limit $S=1/2$. 
 
For the sake of definiteness, we specialize the calculations for the  
pattern of Moriya vectors 
corresponding to the orthorhombic phase in La$_2$CuO$_4$, and give a 
fairly detailed account of the behavior of ground-state energy, anisotropy 
gap, and net ferromagnetic moment. In the last part of this work we 
generalize our approach to describe the geometry of the intermediate phase  
in La$_{2-x}$Nd$_x$CuO$_4$, and discuss the effects of including  
nondegenerate $2p_z$ oxygen orbitals in the calculations. 
\end{abstract} 
 
\newpage 
 
\section{Introduction} 
Dzyaloshinskii\cite{D} was the first to point out that weak  
ferromagnetism in antiferromagnetic compounds can be explained by an  
antisymmetric spin-spin interaction. The microscopic basis for this  
theory was later given by Moriya,\cite{M} who extended Anderson's  
superexchange theory\cite{A} to include the spin-orbit interactions. He 
considered that only the antisymmetric part of the anisotropic spin-spin  
interactions he derived was relevant because it is linear in the spin-orbit  
coupling, while the symmetric anisotropies are of second order. However,  
in a recent work Shekhtman {\it et al.}\cite{S1} showed that the  
symmetric anisotropic term cannot be neglected, and that the whole  
two-spin anisotropic interaction can be mapped onto the isotropic  
Heisenberg interaction by a bond-depending rotation of the original spins.  
This surprising result means that the anisotropy coming from the spin-orbit  
coupling does not lift by itself the ground-state degeneracy, unless the  
required local transformations to map the entire Hamiltonian to the  
isotropic one are not all consistent. Using the mapping to the  
isotropic model these authors showed that infinitely many states with different  
ferromagnetic moments are degenerate with the purely  
antiferromagnetic one. Then, they concluded that some kind of frustration  
of the required local transformations is a necessary ingredient to have a  
definite ferromagnetic moment in the ground-state, a condition that had  
been previously overlooked.   
 
The results of Ref. \cite{S1} are relevant to explain the small  
ferromagnetic  
moment observed in the CuO$_2$ planes of undoped La$_2$CuO$_4$.\cite{Cheong} 
This compound undergoes a structural phase transition from  
tetragonal to orthorhombic symmetry at $T \simeq 500$ K.\cite{Jorgensen} 
In the orthorhombic phase the CuO$_6$ octahedra forming each CuO layer 
tilt in a staggered pattern about the $\langle 110 \rangle$ axis, and 
this distortion combined with the Dzyaloshinskii-Moriya (DM) interactions  
induces a weak ferromagnetic moment in each layer. 
Coffey {\it et al.}\cite{C1,C2} determined the DM  
vectors in both phases from symmetry analysis and detailed microscopic  
calculations. They emphasized that the existence of a net ferromagnetic  
moment requires an alternating pattern of Moriya vectors that change in sign  
from one bond to the next. However, only the consideration of the hidden  
symmetry discussed in \cite{S1} allows one to show that only for the  
orthorhombic symmetries the corresponding pattern leads to the observed small  
ferromagnetic  
moment, while those of the tetragonal phase do not allow it. These results 
where later generalized by Bonesteel\cite{B} to describe the intermediate 
phase observed in La$_{2-x}$Nd$_x$CuO$_4$, where the oxygen octahedra rotate  
around a general axis in the CuO$_2$ plane.\cite{Cr}    
 
In this work we focus on the study of the Heisenberg Hamiltonian with  
DM interaction in the strong quantum limit $S=1/2$. The same Hamiltonian  
was previously studied by semiclassical linear spin-wave theory in Refs. 
\cite{B,S3}, with the emphasis put on magnon dispersion relations.  
However, since the classical ground state is not collinear, relevant  
three-boson terms coming from the Holstein-Primakoff representation of spins  
have been ignored. In this case we extend the Schwinger-boson theory  
discussed in \cite{CGT} to consider the anisotropic DM terms, and  
include magnon-magnon interactions in a saddle-point evaluation of the  
resulting bosonic Hamiltonian. We give a fairly detailed account of the  
energy, ferromagnetic moment, and gap behaviors as a function of the size  
and relative orientation of Moriya vectors, and compare these results  
with similar predictions from linear spin-wave theory. We will first 
specialize the calculations for the  
pattern of microscopic Moriya vectors corresponding to the orthorhombic phase 
in La$_{2}$CuO$_4$. Then, in the last part of this work, we will generalize our 
approach to describe the intermediate phase in  
La$_{2-x}$Nd$_x$CuO$_4$.

The layout of this work is as follows. In Sec. 2 we discuss the  
representation by Schwinger bosons of the anisotropic terms coming from the  
DM interaction. In Sec. 3 we describe the classical ground state of  
the model, and interpret its structure as a spiral order  
in a topologically-equivalent decorated lattice. This allows us to make  
contact with the theory developed in \cite{CGT}. In Sec. 4 we sketch the  
calculation of quantum fluctuations above this classical ground state.  
We give here explicit expressions in terms of the order parameters for the  
quasiparticle dispersion relation, ground-state energy, etc.  
In Sec. 5 we solve numerically the  
consistency equations and present the results for relevant quantities.  
In addition, in Sec. 6 we generalize the calculations to describe the 
intermediate phase in La$_{2-x}$Nd$_x$CuO$_4$ and discuss the effects of  
including the oxygen $2p_z$ orbitals.\cite{S3,Ko} Finally, we end up with  
some conclusions.   
 
\section{Boson representation of the Dzyaloshinskii-Moriya  interaction} 
 
In its simplest derivation,\cite{M} the Heisenberg Hamiltonian including  
the DM interaction can be obtained starting from the  
one-electron Hamiltonian 
 
\begin{equation} 
\label{1e} 
H_{1-el}=\sum_{i,\sigma} \epsilon_i c_{i \sigma}^{\dagger}c_{i \sigma} 
+\sum_{\langle ij \rangle,\sigma \sigma'}  
c_{i \sigma}^{\dagger}(t {\rm e}^{i\theta  
{\bf \widehat{d}}_{ij}.{\bf \sigma} })_ {\sigma \sigma'} c_{j\sigma'}\ , 
\end{equation} 
where $i,j$ are nearest neighbor sites, and the term  
nondiagonal  
in spin space takes into account the spin-orbit interaction. We assume  
nondegenerate ground-state orbitals, and $t$ and $\theta$ are taken  
independent of the bond $\langle ij \rangle$ for simplicity. The unit vector  
${\bf \widehat{d}}_{ij}=-{\bf \widehat{d}}_{ji}$. Then, by using second-order  
perturbation theory in the presence of an on-site 
repulsion $U$\cite{A} one obtains the interaction between spins at sites  
$i,j$ as  
 
\begin{equation} 
\label{E} 
E_{ij}= J {\bf S}_i.{\bf S}_j+ {\bf D}_{ij}^{M}. 
({\bf S}_i \times {\bf S}_j)+{\bf S}_i. {\Gamma}_{ij}. 
{\bf S}_j \ . 
\end{equation} 
In this expression the superexchange interaction $J=(4t^2/U)\cos^2\theta$.  
The last two terms of (\ref{E}) are the   
contributions coming from the spin-orbit interaction of the original  
electrons. They contain the antisymmetric Moriya vector ${\bf  
D}_{ij}^{M}=2J r {\bf \widehat{d}}_{ij} \ (r=\tan\theta)$, and the  
symmetric anisotropy tensor ${\Gamma}$. In  
fact, (\ref{E}) can be most simply written as\cite{S1} 
$$ 
E_{ij}=J \left( {\bf S}_i.{\bf S}_j +  
2r {\bf \widehat{d}}_{ij}.({\bf S}_i \times {\bf S}_j )+ 
r^2 \left[ 2({\bf \widehat{d}}_{ij}.{\bf S}_i) 
({\bf \widehat{d}}_{ij}.{\bf S}_j)- {\bf S}_i.{\bf S}_j \right] \right) 
$$ 
\begin{equation} 
\label{SS} 
=J_0 \ {\bf S}'_i.{\bf S}'_j \ , 
\end{equation} 
where $J_0=4t^2/U$. The spins ${\bf S}'_i,{\bf S}'_j$ are obtained by  
rotating the original ones around the ${\bf \widehat{d}}_{ij}$ axis: 
\begin{equation} 
\label{rot} 
{\bf S}'_k=(1-\cos\theta_k)({\bf \widehat{d}}_{ij}.{\bf S}_k)  
{\bf \widehat{d}}_{ij}+\cos\theta_k \  {\bf S}_k -  
\sin\theta_k \ {\bf S}_k\times {\bf \widehat{d}}_{ij} \ \ \ (k=i,j)\ , 
\end{equation} 
with the rotation angles given by $\theta_i=\theta=-\theta_j$. 
 
At this point we introduce the Schwinger representation ${\bf  
S}'={\bf a}'^{\dagger}. {\bf \sigma} .{\bf a}'$, where the bosonic spinor  
${\bf  
a}'^{\dagger}\equiv (a'^{\dagger}_{\uparrow},a'^{\dagger}_{\downarrow})$.  
Then, the rotational invariant structure in (\ref{SS}) can be written in  
the usual way\cite{CGT} as   
\begin{equation} 
\label{ab} 
{\bf S}'_i.{\bf  
S}'_j=:B'^{\dagger}_{ij}B'_{ij}:-A'^{\dagger}_{ij}A'_{ij}\ , 
\end{equation}  
in terms of the $SU(2)$ singlets  
$$ 
B'^{\dagger}_{ij}={\frac 1 2}\sum_{\sigma} a'^{\dagger}_{i \sigma} a'_{j  
\sigma} \ \ , \ \   
A'_{ij}={\frac 1 2}\sum_{\sigma} \sigma a'_{i \sigma} a'_{j, -\sigma} \ . 
$$
In (\ref{ab}) the colons denote a normal-ordered product of Bose 
operators.  

From standard rotation properties we have: 
$$ 
{\bf S}'={\cal R}_{\theta}{\bf S}={\bf a}^{\dagger}.({\cal R}_{\theta}  
{\bf \sigma}) .{\bf a}=({\bf a}^{\dagger}{\cal  
U}^{\dagger}_{\theta}).{\bf \sigma} .({\cal U}_{\theta} 
{\bf a})={\bf a'^{\dagger}. \sigma .a'} \ , 
$$ 
where ${\cal R}_{\theta}$ is the three-dimensional matrix that  
rotates an angle $\theta$ around the ${\bf \widehat{d}}_{ij}$ axis, and  
${\cal U}_{\theta}=\exp[-i(\theta/2) {\bf \widehat{d}}_{ij}.{\bf \sigma}]$ is  
the corresponding $SU(2)$ matrix. In this way we have   
${\bf a'}={\cal U}_{\theta}{\bf a}$, and this relation can be replaced in  
the definitions of the $A',B'$ singlets in order to express them in terms  
of the unrotated spinors:   
\begin{equation} 
\label{abcd} 
B'^{\dagger}_{ij}=\cos\theta B^{\dagger}_{ij}+ 
\sin\theta C^{\dagger}_{ij} \ \ \ ,\ \ \    
A'_{ij}=\cos\theta A_{ij}-\sin\theta D_{ij}.   
\end{equation} 
Here, in addition to the singlets $A,B$ we have defined 
$$  
C^{\dagger}_{ij}={\frac 1 2}\sum_{\sigma \sigma'}  
a^{\dagger}_{i\sigma}(i{\bf  
\widehat{d}}_{ij}. {\bf \sigma})_{\sigma \sigma'}a_{j\sigma'} \ \ , \ \  
D_{ij}={\frac 1 2}\sum_{\sigma,\sigma'} a_{i\sigma} 
(\sigma^{y}{\bf \widehat{d}}_{ij}.{\bf \sigma})_{\sigma \sigma'}  
a_{j\sigma'} \ . 
$$ 
By replacing (\ref{ab}),(\ref{abcd}) in (\ref{SS}) we can identify  
the correct expressions in terms of Schwinger  
bosons of the spin interactions: 
$$ 
{\bf S}_i.{\bf S}_j=:B^{\dagger}_{ij}B_{ij}:-A^{\dagger}_{ij}A_{ij} 
$$ 
\begin{equation} 
\label{rep} 
{\bf \widehat{d}}_{ij}.({\bf S}_i  
\times {\bf S}_j)= {\frac 1 2} (:B^{\dagger}_{ij} 
C_{ij}+C^{\dagger}_{ij}B_{ij}:+A^{\dagger}_{ij}D_{ij}+ 
D^{\dagger}_{ij}A_{ij}) 
\end{equation} 
$$ 
2 ({\bf \widehat{d}}_{ij}.{\bf S}_i) ({\bf \widehat{d}}_{ij}.{\bf S}_j)- 
{\bf S}_i.{\bf S}_j = :C^{\dagger}_{ij}C_{ij}:-D^{\dagger}_{ij}D_{ij}\ . 
$$ 
In this way the DM terms are given as products of  
bilinear boson operators that have the same rotational properties of the  
original spin interactions. This allows us to perform a conserving  
mean-field approximation by breaking these products. However, before  
undertaking the study of the resulting mean-field Hamiltonian, in the  
next section we specialize the above results for the CuO$_2$ planes  
in the low-temperature orthorhombic (LTO) phase of La$_2$CuO$_4$, and  
discuss how to describe the structure of the corresponding classical  
ground states as commensurate spiral orders. 
 
\section{The classical ground state as a spiral order} 
 
If the rotations required at every site to bring the spin-spin interactions  
for {\it all} bonds in the lattice to the form (\ref{SS}) are compatible,  
then the Hamiltonian $H_{spin}=(1/2)\sum_{\langle ij \rangle}  
E_{ij}$ is isomorphic to the  
isotropic Heisenberg model. This happens when the product of the 
four rotations around each basic plaquette in the lattice equals unity,  
which corresponds to the {\it unfrustrated} case. In this case, in  
correspondence with the ground-state degeneracy of (\ref{SS})  
there is also a large degeneracy in the ground state of $H_{spin}$.  
>From the relations (\ref{rot}) between original and rotated spins 
we see that by judiciously chosing the direction of the N\'eel we can  
find ground states of  
$H_{spin}$ with no net ferromagnetic moment (${\bf S}'_i=-{\bf S}'_j$  
parallel to ${\bf \widehat{d}}_{ij}$) up to a maximum ferromagnetic moment  
of $\pm \sin\theta$\ (${\bf S}'_i=-{\bf S}'_j$ perpendicular to  
${\bf \widehat{d}}_{ij}$). Even though ${\bf D}^M_{ij}$ in (\ref{E}) points  
in a special direction in spin space, because of the presence of the  
symmetric tensor ${\Gamma}$ this anisotropy does  
not lift the degeneracy of the ground state.\cite{S1} 
 
According to the above discussion, in order to have a net ferromagnetic  
moment there should be some degree of {\it frustration} in the  
system.\cite{S1}  
That is, there should exist no spin rotations that map the entire  
Hamiltonian onto an isotropic one. In particular, in the LTO phase of  
La$_2$CuO$_4$ the crystal structure leads to 
the pattern of Moriya vectors shown in Fig. 1. In this figure  
${\bf D}_1=D{\bf \widehat{d}}_1$ 
and ${\bf D}_2=D{\bf \widehat{d}}_2$, where ${\bf \widehat{d}}_1= 
(\sin\beta,\cos\beta,0)$,\   
${\bf \widehat{d}}_2=(\cos\beta,\sin\beta,0)$, and $D=2Jr$ as in the  
previous section\cite{Nota}. Then, the magnetic structure can be  
described in terms of two interpenetrating sublattices. Defining ${\bf  
D}^{\pm} =({\bf D}_1 \pm {\bf D}_2)/2$, it has been shown\cite{S2} that the  
system  
develops a net ferromagnetic moment only when both ${\bf D}^{-}$ and ${\bf  
D}^{+}$ are different from zero $(\beta \neq \pm \pi/4)$. In  
such a case the staggered magnetization of the ground state is directed  
along ${\bf D}^{-}$, and the ferromagnetic moment is proportional to  
${\bf D}^{+} \times {\bf D}^{-}$ (see Fig. 2). 
 
In the two-sublattice picture the classical ground state has magnetizations  
${\bf S}_A=(S\cos\phi,0,S\sin\phi)$ and ${\bf  
S}_B=(-S\cos\phi,0,S\sin\phi)$, with $\tan\phi=r \cos(\beta-\pi/4)$  
(We are using the $x$-axis in spin space directed along ${\bf D}^-$ and 
the $z$-axis perpendicular to the CuO$_2$ plane). In  
order  
to make contact with the theory developed in \cite{CGT} it is convenient  
to describe this state as a commensurate spiral order in a distorted  
lattice. To this end we move the sites belonging to sublattices $A,B$ by   
vectors ${\vec \delta}_A=(\phi /\pi,0)$ and ${\vec \delta}_B=-{\vec \delta}_A$ 
respectively (see Fig. 3), so that the  
basis vectors of the new decorated square lattice are ${\vec  
\delta}_A$ and ${\vec \delta}_1=(1,0)+{\vec \delta}_B$ \  
(We took the original lattice constant $a=1$  
and used arrows instead of boldface to indicate vectors in real space).  
The magnetic order is then given by  
${\bf S}_{\vec r}=(S\cos {\vec Q}.{\vec r},0,S\sin {\vec Q}.{\vec r})$,  
where the magnetic wavevector 
${\vec Q}=(\pi,\pi)$ and the ${\vec r}$'s indicate the sites of the  
distorted lattice. This classical ground state can be  
described by condensing the Schwinger bosons according to  
$a^{\dagger}_{{\vec r}\sigma}=a_{{\vec r}\sigma}=\sqrt{S}(\cos{\vec  
Q}.{\vec r}/2 +\sigma \sin{\vec Q}.{\vec r}/2)$. In the next  
section we will incorporate  
the quantum fluctuations on this description by considering the  
Schwinger-boson dynamics to mean-field order. 
 
\section{Quantum description to mean-field order } 
 
Once $H_{spin}$ is expressed in terms of Schwinger bosons as indicated in  
Sec. 2 (see (\ref{rep})), all the four-boson products are decoupled using  
as order parameters the mean values of the  
correlations $A,B,C,D$\ between sites in different sublattices. Then, the  
mean-field Hamiltonian becomes: 
\begin{equation} 
\label{MFH} 
H_{\rm MF}=\sum_{i,{\vec \delta}} [ {\cal B}({\vec \delta})  
{\cal B}_{i,i+{\vec 
\delta}}^{\dagger}-{\cal A}({\vec \delta}) {\cal A}^{\dagger}_{i,i+{\vec  
\delta}} + {\rm H.c.} ] - N\sum_{\vec \delta} (|{\cal B}({\vec  
\delta})|^2 - |{\cal A}({\vec \delta})|^2 ), 
\end{equation} 
where ${\cal B}_{i,i+{\vec \delta}}\equiv B_{i,i+{\vec \delta}}+r  
C_{i,i+{\vec \delta}}\ , \  {\cal A}_{i,i+{\vec \delta}}\equiv 
A_{i,i+{\vec \delta}}-r D_{i,i+{\vec \delta}}$\ , and  
${\cal B}({\vec \delta}), {\cal A}({\vec \delta})$ are their  
corresponding mean-field values. Here $N$ is the number of decorated unit  
cells and $i$ runs on the $A$ sublattice. In particular, because of the  
geometry of Moriya vectors in Fig. 1, the $C,D$ operators take the simple  
form 
$$ 
C^{\dagger}_{i,i+{\vec \delta}}={\frac 1 2}\sum_{\sigma} \sigma  
{\rm e}^{i\sigma \eta({\vec \delta})} a^{\dagger}_{i\sigma}  
a_{i+{\vec \delta},-\sigma} \ \ , \ \ D_{i,i+{\vec  
\delta}}={\frac 1 2}\sum_{\sigma}   
{\rm e}^{-i\sigma \eta({\vec \delta})} a_{i\sigma}  
a_{i+{\vec \delta},\sigma}\ , 
$$  
with $\eta({\vec \delta}_j)=(-)^j(\pi/4-\beta)$ \ \ ($j=1,4$).  
 
We perform now a transformation to reciprocal space according to 
$$ 
a_{i \sigma}={\frac {1} {\sqrt{N}}}\sum_{\vec k} \alpha_{{\vec k} \sigma} 
{\rm e}^{-i{\vec k}_{\sigma}.{\vec r}_i} \ \ , \ \   
a_{i+{\vec \delta}, \sigma}={\frac {1} {\sqrt{N}}}\sum_{\vec k}  
\beta_{{\vec k} 
\sigma} {\rm e}^{-i{\vec k}_{\sigma}.{\vec r}_{i+{\vec \delta}}}\ , 
$$ 
where ${\vec k}_{\sigma}={\vec k}+\sigma {\vec Q}/2$ and the  
${\vec k}$'s are the normal modes of the decorated lattice. The  
resulting Hamiltonian has to be supplemented by adding, through Lagrange 
multipliers $\lambda_A,\lambda_B$,   
the restrictions on the number of bosons on each sublattice: 
$$ 
\sum_{{\vec k} \sigma} \langle \alpha^{\dagger}_{{\vec k}\sigma} 
\alpha_{{\vec k}\sigma}\rangle =2SN \ \ \ , \ \ \   
\sum_{{\vec k} \sigma} \langle \beta^{\dagger}_{{\vec k}\sigma} 
\beta_{{\vec k}\sigma}\rangle =2SN \ . 
$$ 
Notice, however, that these restrictions are only the average of the exact  
constraints {\it per 
site} required to have a faithful representation of the spin algebra.  
With this simplication we obtain a $8 \times 8$ dynamical matrix  
which has to be diagonalized numerically. However, the solution of the  
mean-field equations for the ${\cal A,B}$ order parameters shows that,  
for the 
symmetry of the spiral order, the parameters ${\cal B}({\vec \delta})=0$  
and $\lambda_A=\lambda_B$. Then, the original dynamical matrix decouples in  
two $4 \times 4$ blocks given by: 
\begin{equation} 
\label{delta} 
\Delta({\vec q})=\left(\matrix{ 
\lambda  &0   &-\gamma({\vec q})  &\gamma_{+}({\vec q})\cr 
0  &\lambda  &\gamma_{-}({\vec q})  &\gamma({\vec q})\cr   
-{\gamma}^{*}({\vec q})  &{\gamma}_{-}^{*}({\vec q})  &\lambda &0\cr    
{\gamma}_{+}^{*}({\vec q})  &{\gamma}^{*}({\vec q})  &0  &\lambda 
\cr} \right)  
\end{equation} 
and the corresponding one with ${\vec q}\equiv {\vec k}+{\vec  
Q}/2$ replaced by $-{\vec q}$\ . We have defined: 
$$ 
\gamma ({\vec q})={\frac 1 2}\sum_{\vec \delta} {\cal A}({\vec \delta}) 
{\rm e}^{-i{\vec q}.({\vec \delta}-{\vec \delta}_A)} \ \ , \ \  
\gamma_{\pm}({\vec q})={\frac 1 2}\sum_{\vec \delta} r {\cal A} 
({\vec \delta}){\rm e}^{-i{\vec q}.({\vec \delta}-{\vec \delta}_A) 
\pm i\eta({\vec \delta})} \ .  
$$ 
Paraunitary diagonalization of (\ref{delta}) produces the  
quasiparticle dispersion relations 
$$ 
\omega^2_{\pm}=\lambda^2-\left( |\gamma|^2+{\frac {1}  
{2}}(|\gamma_{+}|^2+|\gamma_{-}|^2) \right) \pm   
\sqrt{ {\frac {1} 4} \left( |\gamma_{+}|^2-|\gamma_{-}|^2 \right)^2  
+ |\gamma^{*}\gamma_{+}- \gamma \gamma_{-}^{*}|^2 } \ . 
$$ 
Numerical evaluation (see next section) shows that the in-plane and  
out-of-plane gaps are degenerate.  
For the unfrustrated case $\beta=\pi/4$, so that $\gamma_{\pm}=\gamma$,  
and this equation reduces to  
$\omega^2_{\pm}=\lambda^2-(1+r^2)|\gamma|^2$. As expected, these  
degenerate quasiparticle bands correspond to an antiferromagnet with  
renormalized interaction $J(1+r^2)=J/\cos^2\theta \equiv J_0$ \ (see 
(\ref{SS})).  
 
The ground-state energy is given by  
$E_0=(1/2)\sum_{{\vec k} \sigma}\omega_{\sigma}({\vec  
k})-(2S+1)N\lambda$\ . The mean-field equations that determine the  
parameters $\lambda$ and ${\cal A}({\vec \delta})$  
are obtained by minimization of $E_0$. In the next section we discuss the  
results obtained by solving numerically these equations.  
 
\section{Results for the LTO phase of La$_2$CuO$_4$} 
 
We have solved numerically the consistency equations by iteration,  
starting from the classical values of the order parameters and Lagrange  
multiplier (The classical values of the order parameters can be obtained  
by using the structure of the boson condensate given at the end of Sec. 3,  
and $\lambda_{cl}=-E_{cl}/NS^2=2(1+r^2)$). For the  
unfrustrated case $\beta=\pi/4$\  we obtain, as expected, two doubly-degenerate 
quasiparticles branches corresponding to an isotropic antiferromagnet  
with the effective coupling $J_0$. For $\beta$ infinitesimally smaller than  
$\pi/4$ one of these branches develops gaps at  
${\vec k}={\vec 0}$, which correspond to the degenerate  
in- and out-of-plane gaps  
associated to the anisotropy. The other branch remains   
very much unaffected, i.e., almost identical to the isotropic renormalized 
antiferromagnetic dispersion. For large $S$ the 
gapped branch becomes the linear spin-wave dispersion  
relation.\cite{B,S3} In Fig. 4 we plot both branches for $\beta=0$ and 
$r=0.4$\  along the  
typical $(\Gamma,X,M)$ path there indicated. The dashed line  
gives the results of linear spin-wave theory for comparison.  
The fact that one of the branches remain gapless can be understood by  
noticing that in the Schwinger-boson approach the existence of a local  
magnetization is intimately linked to the boson condensation. Then, in  
order for the system to have a magnetized ground state, the chemical  
potential (Lagrange multiplier) must be pinned to the bottom of one of  
the original boson bands, giving a gapless quasiparticle branch. This  
can be confirmed by simply studying the isotropic antiferromagnet with a  
staggered magnetic field. On the other hand, from a  
physical point of view this is not a problem since the exact form of the  
constraint requires creation and destruction of both kind of  
quasiparticles to describe a magnon. In Fig. 5 we plot the gap as a  
function of $r$ for $\beta=0$\cite{Nota1}. 
Notice that for the physical value $r \simeq 0.1$ the result  
is indistinguishable from the spin wave prediction. 
 
The classical energy is given by $E_{cl}=-2(1+r^2)$, and is  
independent of the frustration. In Fig. 6 we plot the ratio between the  
quantum  
and classical ground-state energies as a function of $\beta$ for several  
values of $r$. As can be seen, quantum fluctuations are sensitive to the  
degree of frustration, although the changes in energy are not important  
even for  
unphysically large values of $r$. In Fig. 7 we plot the behavior of the  
net ferromagnetic moment for the same values of $r$. As  
expected,\cite{S2} it goes to zero for $\beta=-\pi/4$, where the  
Dzyaloshinskii vector ${\bf D}^{+}$ vanishes and the  
ground-state configuration is completely antiferromagnetic (see Sec. 3).     
The values at $\beta=\pi/4$ correspond to the ferromagnetic moments  
in the unfrustrated case (${\bf D}^{-}=0$). However, these moments are  
unobservable since, as pointed out above, in this case the Hamiltonian  
becomes isomorphic to the isotropic Heisenberg model.\cite{S1} Notice that 
this happens only exactly at $\beta=\pi/4$; for slightly different values 
the system develops a relatively large net ferromagnetic moment. Notably, as 
shown in Fig. 7 in the quantum case the maximum ferromagnetic moment is 
obtained for some intermediate value of $\beta$ which depends weakly on 
$r$. 
 
\section{Intermediate phase in La$_{2-x}$Nd$_x$CuO$_4$} 
 
In the LTO phase the CuO$_6$ octahedra forming each CuO layer tilt in 
a staggered pattern about the $\langle 110 \rangle$ axis. In this section 
we consider the effects of DM interactions which occur in the presence 
of tilting distorsions about an arbitrary axis. Such general tilting 
distorsions may have physical relevance for La$_{2-x}$Nd$_x$CuO$_4$.\cite{B} 
 
In order to generalize the above calculations for an arbitrary  
direction of the octahedra tilting axis we have to perform the 
following changes: i) In Sec. 2, the parameters 
$t$ and $\theta$ of (\ref{1e}) must be bond dependent, which implies that 
$|D^M_{ij}|$ is also bond dependent; ii) In Sec. 3, the new pattern 
of Moriya vectors is given by ${\bf D}_1=(\lambda_1\cos \alpha, 
\lambda_2 \sin\alpha,0)$ and ${\bf D}_2=(\lambda_2\cos \alpha, 
\lambda_1 \sin\alpha,0)$, with $\alpha$ the angle between the octahedra  
rotation axis and the $[100]$ direction in the CuO plane, and $\lambda_1 
\simeq 0$\cite{B} (The angle $\alpha$ equals zero in the tetragonal  
phase and $\pi /4$ in the orthorhombic phase). Furthermore, the angle $\phi$ 
which characterizes the classical ground-state spin configuration  
satisfies $\sin \phi =\sin \psi \sin 2\alpha$, where $\tan \psi=\lambda_2/ 
2J$\cite{S3}; iv) In Sec. 4, the parameter $r=|D^M_{ij}|/2J$ is now 
bond dependent. This leads to different ${\cal A}({\vec \delta})$ for 
each bond direction, but ${\cal B}({\vec \delta})$ remains zero and 
the dispersion relations have the same formal expression. With these changes 
we obtained the following results: The ground-state energy and gap depend 
slightly on the tilting axis angle $\alpha$. On the contrary, the net 
ferromagnetic moment is strongly dependent on $\alpha$ (see Fig. 8). It 
vanishes in the tetragonal phase ($\alpha=0$) and reaches its maximum value 
in the LTO phase ($\alpha=\pi/4$). 
 
In \cite{Ko} it has been suggested that the inclusion of the oxygen $2p_z$  
orbitals would account for the weak ferromagnetism in the LTO phase of 
La$_2$CuO$_4$, with the frustration origin advocated in \cite{S1} 
being irrelevant. However, 
Entin-Wohlman {\it et al.}\cite{S3} showed that the effects of introducing 
these orbitals is too small to be responsible for the observed weak 
ferromagnetism. On the other hand, the multilevel mechanism does break the  
degeneracy of the in- and out-of-plane gaps, although the relative  
magnitude of this breaking is again too small when compared to experiments. 
We have also considered including the $2p_z$ orbitals in our  
calculations. In this case the required changes are the following: i) In 
Sec. 2, Eq. (\ref{SS}) has to be replaced by 
$$ 
E_{ij}=J {\bf S}_i.{\bf S}_j +  
{\bf D}_{ij}.({\bf S}_i \times {\bf S}_j )+ 
\frac {1} {4J} (1-\Omega) \left[ 2({\bf D}_{ij}.{\bf S}_i) 
({\bf D}_{ij}.{\bf S}_j)- |D_{ij}|^2 {\bf S}_i.{\bf S}_j \right] \ , 
$$ 
where $\Omega$ is proportional to the energy difference between the 
$2p_z$ and $2p_\sigma$ orbitals of the oxygen\cite{S3}; ii) In Sec. 3, 
the angle $\phi$ that describes the classical ground state has a complex 
expression in terms of the parameters of the model. We have obtained 
it by numerical minimization of the classical energy; iii) In Sec. 4, the  
mean-field 
Hamiltonian (\ref{MFH}) has to be written as a function of the four 
order parameters $A,B,C$ and $D$. The $8\times 8$ dynamical matrix no longer 
decouples in two blocks, so that its paraunitary diagonalization has to 
be performed numerically. In Fig. 9 we present the new dispersion relations    
obtained in this case, for $\lambda_1=0\ ,\lambda_2=0.6\ ,\alpha=\pi/8$\ , and 
$\Omega=0.4$. As can be seen, the degenerate gapped branch splits into two 
branches that differ mostly near the gap wavevector. However, even for 
this large $\Omega$ the ratio  
between the in- and out-of-plane gaps is still much smaller than the  
experimental value. Finally, in Fig. 10 we plot the ferromagnetic 
moment for different values of $\Omega$. Remarkably, for some  
special value of $\alpha$ this moment becomes independent of $\Omega$.

\section{Conclusions} 
 
We presented a Schwinger-boson approach to the Heisenberg  
model with DM interaction. We showed how to represent the anisotropic  
DM interactions in terms of Schwinger bosons keeping the correct  
symmetries present in the spin representation, which allowed us to perform  
a conserving mean-field approximation. Unlike previous studies of this  
model by  
linear spin-wave theory, our approach takes into account magnon-magnon  
interactions already at mean-field order and includes the effects of  
three-boson terms, characteristic of noncollinear phases, which are  
ignored in standard studies. Our results recover the linear spin-wave  
predictions in the semiclassical large-$S$ limit, and show a small  
renormalization in the strong quantum limit $S=1/2$.  
 
Although the emphasis was put on the basic properties of the model 
Hamiltonian studied, for definiteness we considered a pattern of 
Moriya vectors corresponding to the LTO phase in La$_2$CuO$_4$. In the 
last section we generalized the calculations to discuss the behavior of  
relevant quantities for arbitrary directions 
of the octahedra tilting axis, which might be relevant 
to describe the intermediate phase in La$_{2-x}$Nd$_x$CuO$_4$. Finally, we  
have also considered the inclusion of nondegenerate $2p_z$ oxygen 
orbitals, and showed that they cannot account for the experimentally observed 
difference between in- and out-of-plane gaps.

\begin{figure}[ht]
\begin{center}
\epsfysize=13cm
\leavevmode
\epsffile{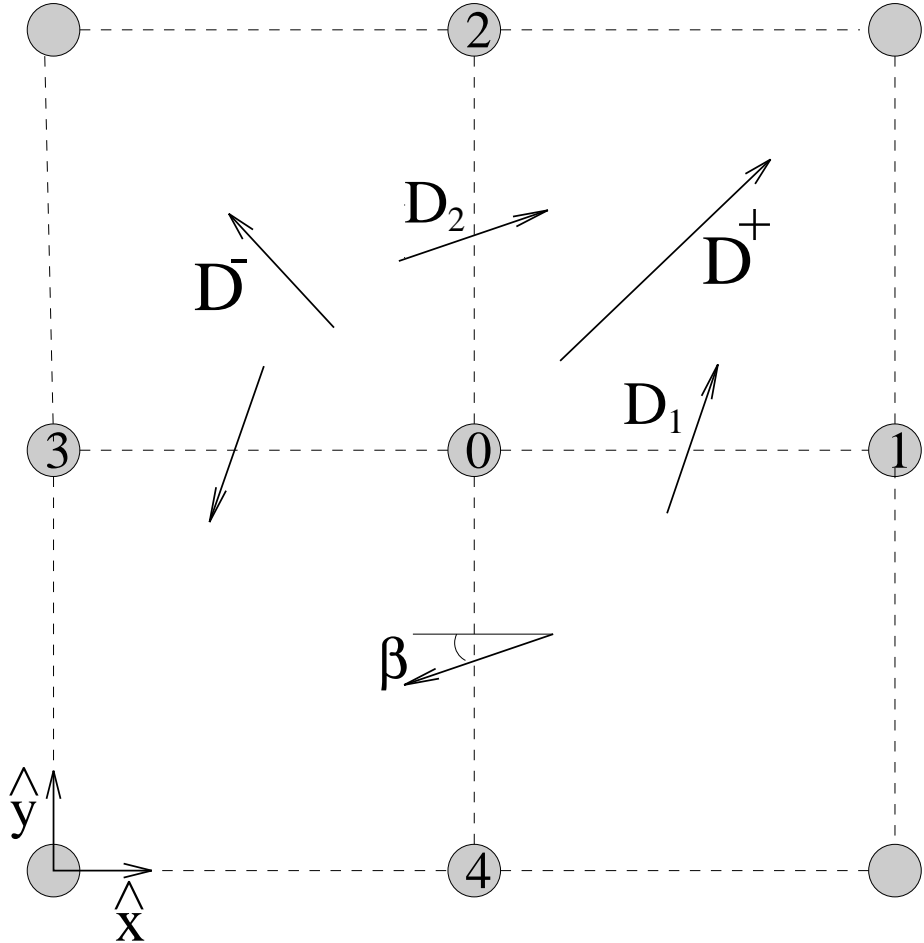}
\caption
{Copper sites of the CuO$_2$ planes in La$_2$CuO$_4$. We indicate  
the bond-depending Moriya vectors ${\bf D}^M_{ij}$, which 
alternate in sign along the $x$ and $y$ axis. Also shown are the corresponding  
${\bf D}^{\pm}$ vectors defined in the main text. } 
\end{center}
\end{figure}

\begin{figure}[ht]
\begin{center}
\epsfysize=13cm
\leavevmode
\epsffile{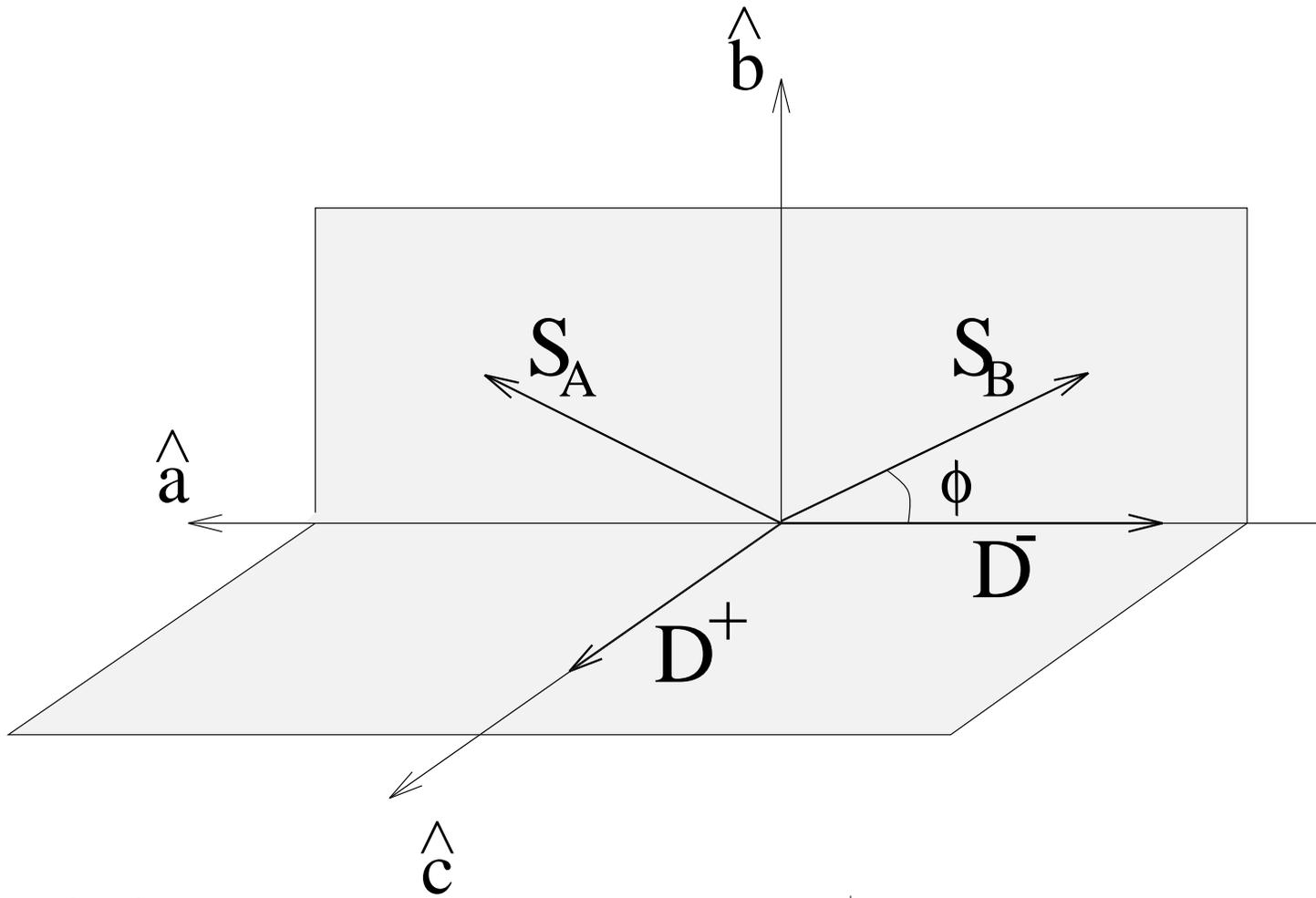}
\caption
{Sublattice magnetizations  
${\bf S}_A,{\bf S}_B$ and their relations to the ${\bf D}^{\pm}$ vectors and 
orthorhombic axis.}  
\end{center}
\end{figure}

\begin{figure}[ht]
\begin{center}
\epsfysize=13cm
\leavevmode
\epsffile{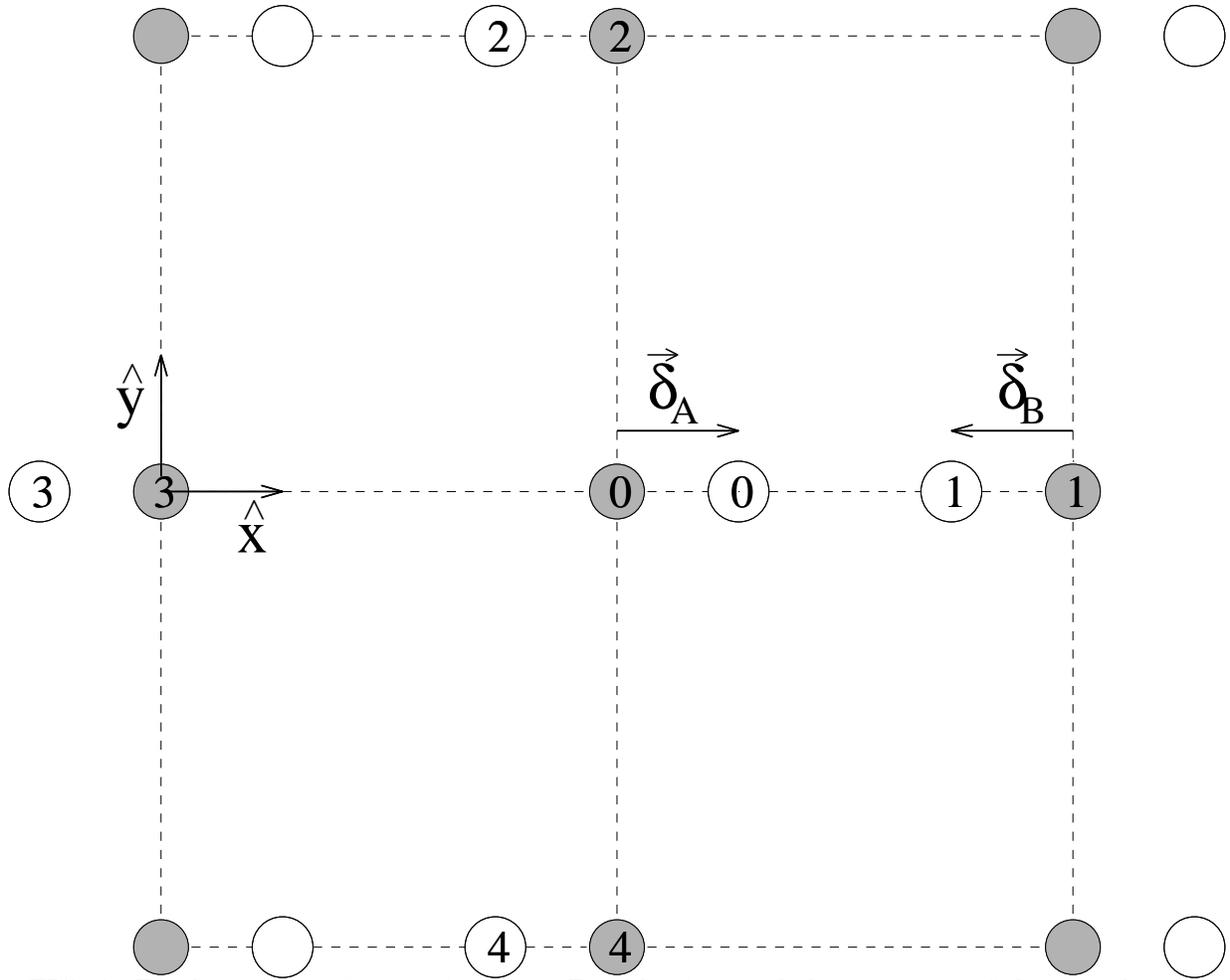}
\caption
{Displacements of atoms in the $A,B$ sublattices and the structure of  
the new decorated square lattice.} 
\end{center}
\end{figure}

\begin{figure}[ht]
\begin{center}
\epsfysize=16cm
\leavevmode
\epsffile{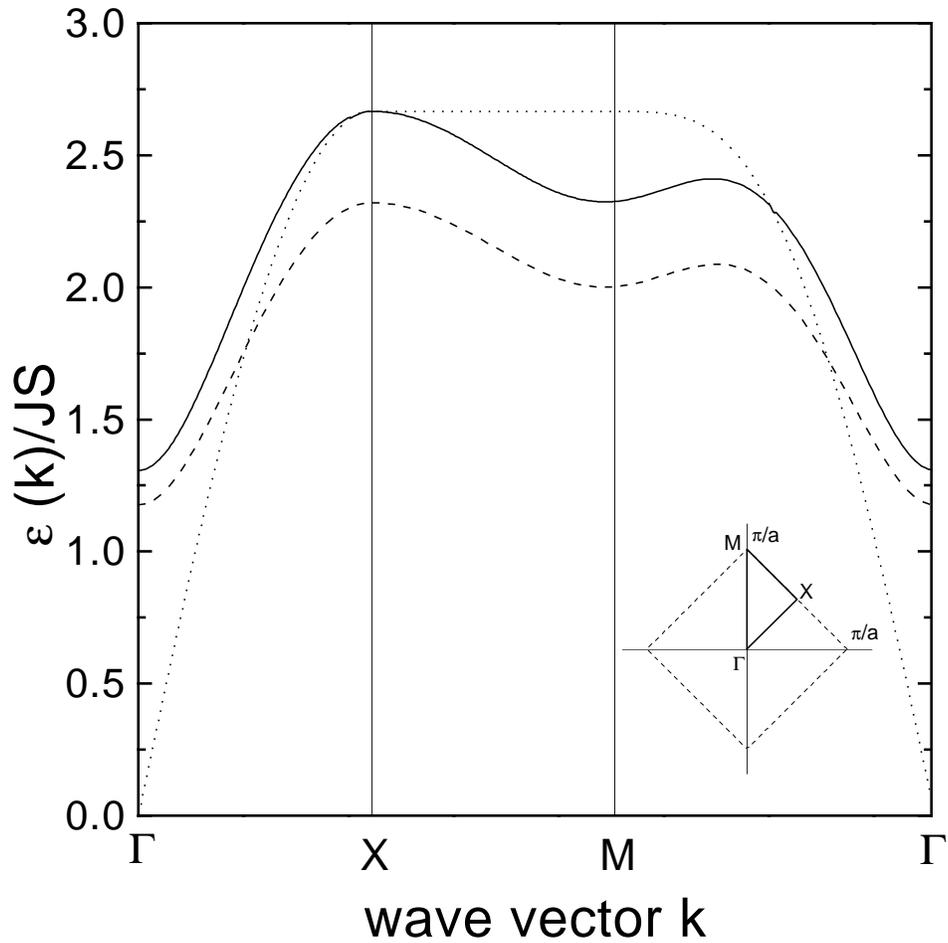}
\caption
{Quasiparticle dispersion relations for $r=0.4$ and $\beta=0$,  
along the $\Gamma-X-M$ path shown in the inset. The full line is the 
Schwinger-boson result for the degenerate gapped branch; the dashed line 
indicates the linear spin-wave prediction for comparison. 
The dotted line gives the degenerate gapless branch discussed in the main 
text. } 
\end{center}
\end{figure}

\begin{figure}[ht]
\begin{center}
\epsfysize=16cm
\leavevmode
\epsffile{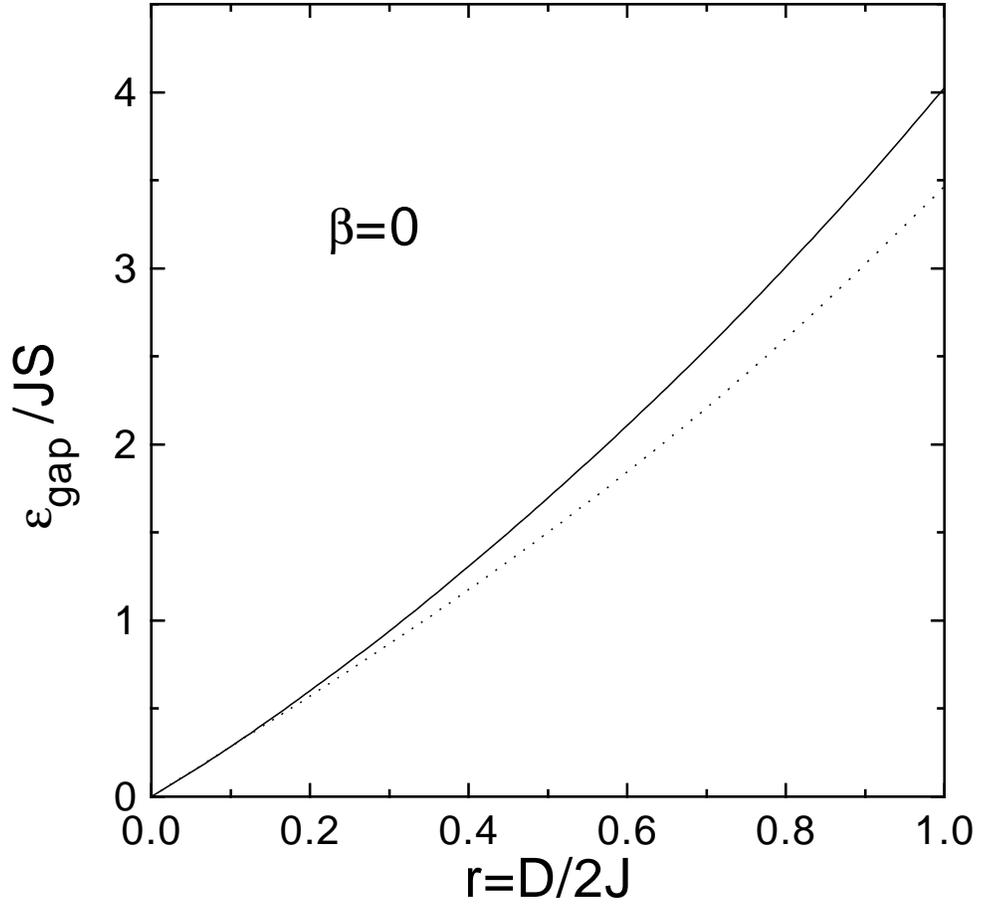}
\caption
{Anisotropy gap as a function of $r$ for $\beta=0$. The full line 
is the Schwinger-boson prediction; the point line gives the linear spin-wave 
result for comparison.} 
\end{center}
\end{figure}

\begin{figure}[ht]
\begin{center}
\epsfysize=20cm
\leavevmode
\epsffile{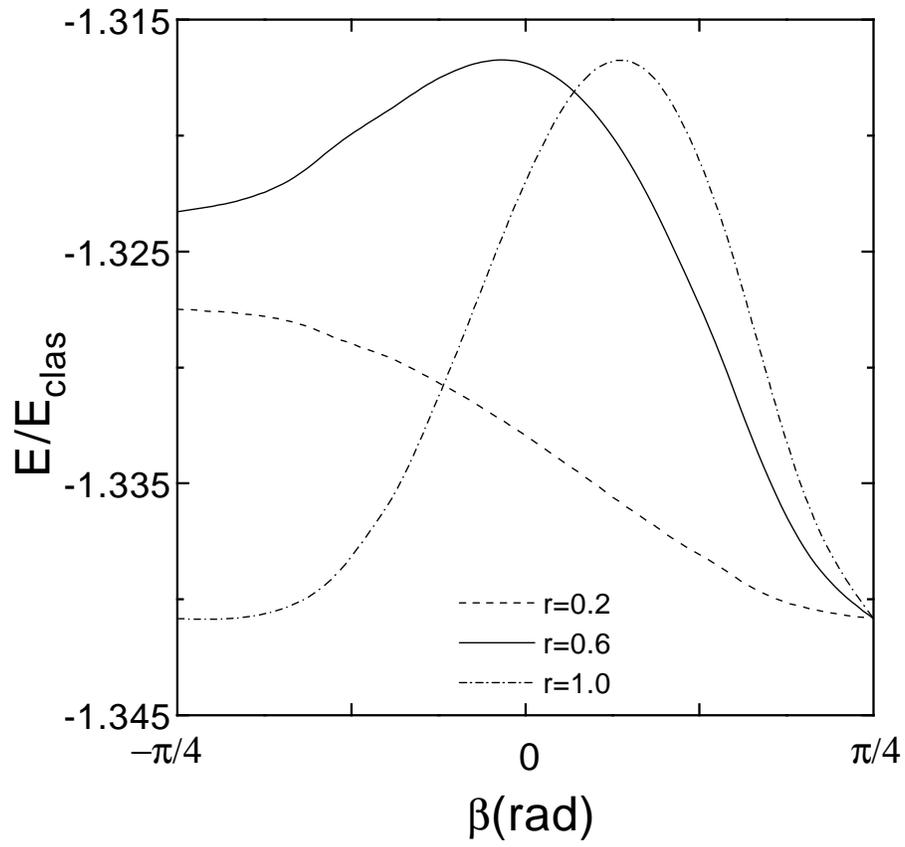}
\caption
{Ratio between classical and quantum ground-state energies as a  
function of $\beta$ for different values of $r$.} 
\end{center}
\end{figure}

\begin{figure}[ht]
\begin{center}
\epsfysize=16cm
\leavevmode
\epsffile{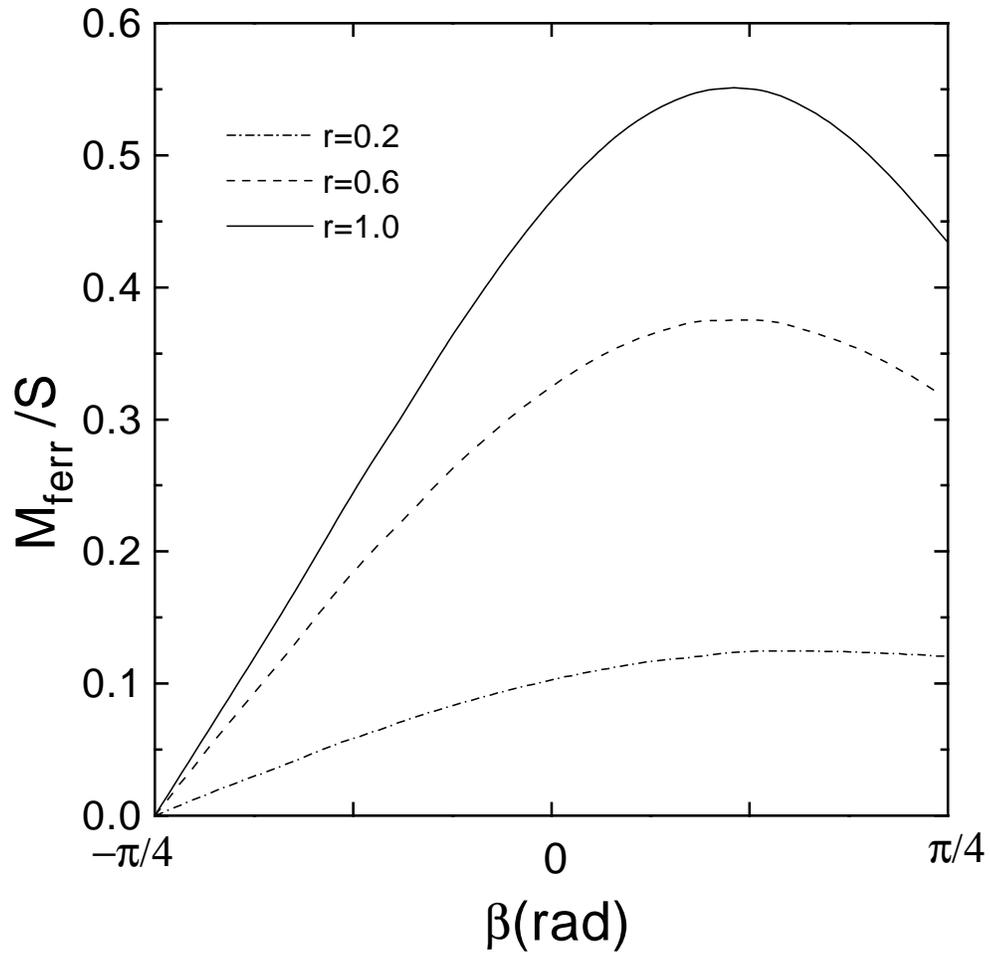}
\caption
{Net ferromagnetic moment as a function of $\beta$ for different 
values of $r$.} 
\end{center}
\end{figure}

\begin{figure}[ht]
\begin{center}
\epsfysize=16cm
\leavevmode
\epsffile{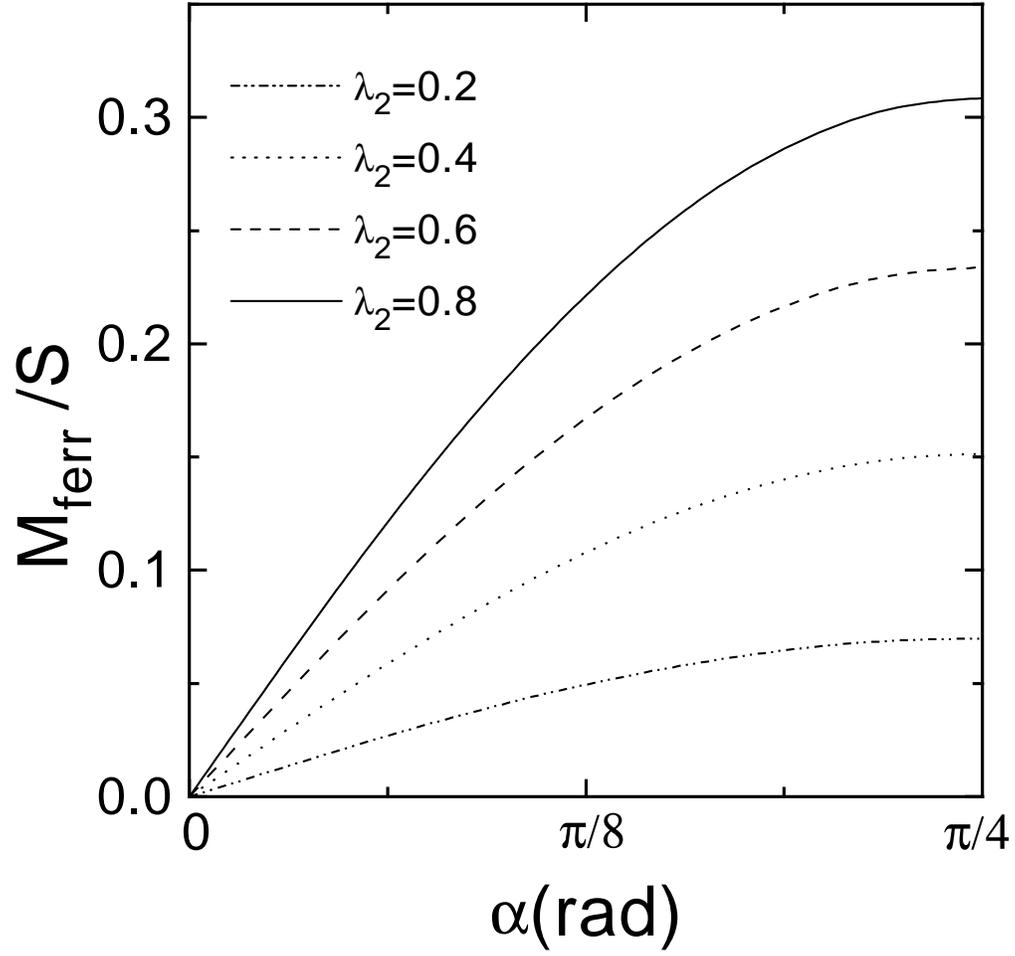}
\caption
{Net ferromagnetic moment as a function of the direction of the 
octahedra tilting axis. See the main text for the definitions of the
parameters involved. } 
\end{center}
\end{figure}

\begin{figure}[ht]
\begin{center}
\epsfysize=16cm
\leavevmode
\epsffile{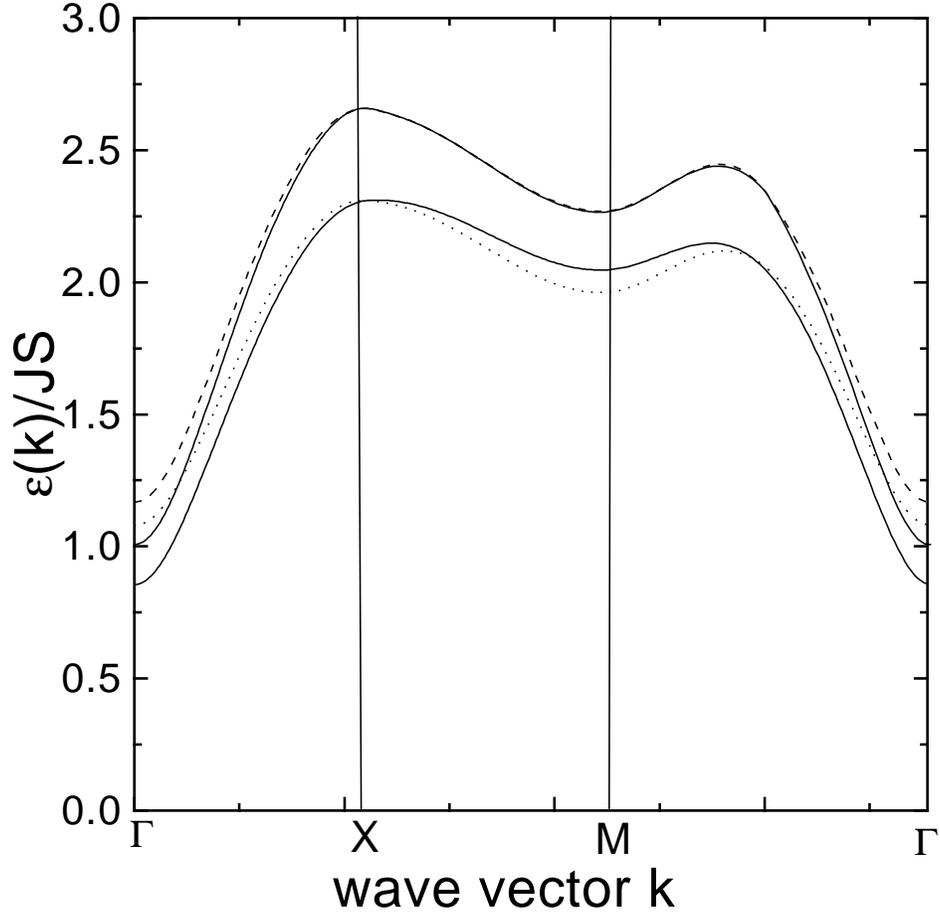}
\caption
{Splitting of the degenerate gapped quasiparticle branches for 
$\Omega=0.4$. Upper full and dashed lines: Schwinger-boson results. 
Lower full and point lines: linear spin-wave predictions.} 
\end{center}
\end{figure}

\begin{figure}[ht]
\begin{center}
\epsfysize=20cm
\leavevmode
\epsffile{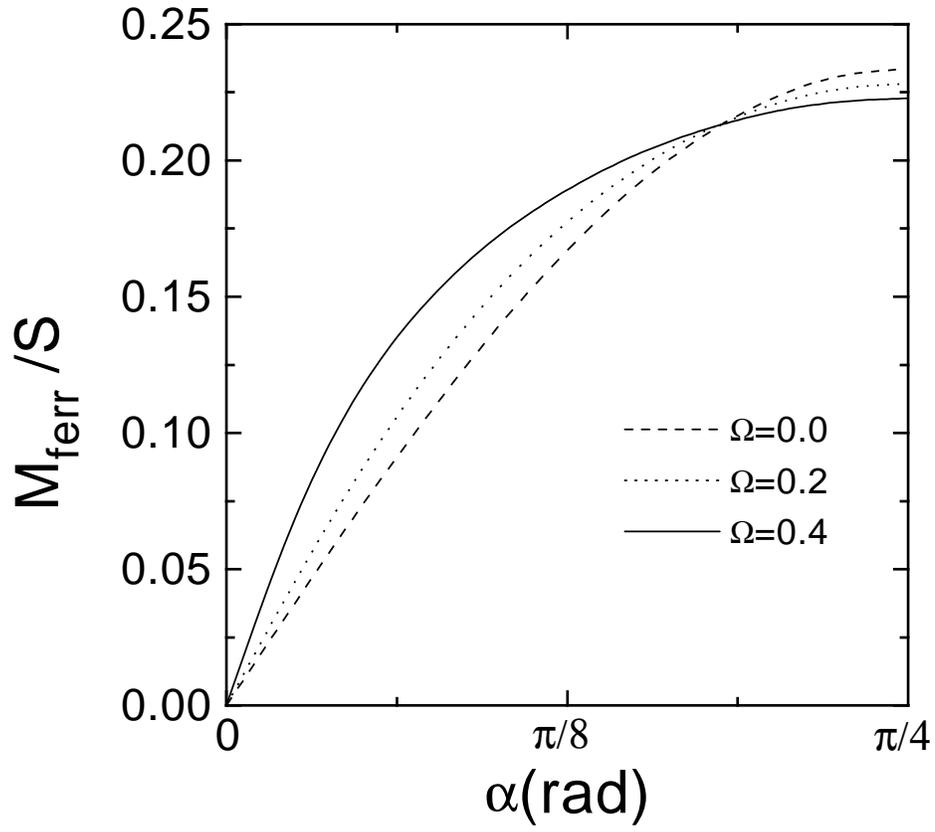}
\caption
{Net ferromagnetic moment as a function of the direction of the 
octahedra tilting axis for different values of $\Omega$.} 
\end{center}
\end{figure}


\begin{thebibliography}{99}
\bibitem{Nota} For generality we will consider the behavior of relevant  
quantities for arbitrary values of $\beta$, although it is in fact  
almost zero\cite{C2}. Notice that because of the symmetries  
it is only necessary to consider $-\pi/4 \leq \beta \leq \pi/4$.
\bibitem{Nota1} The physical values in the LTO phase of La$_2$CuO$_4$ are  
$\beta \simeq 0$ and $r \simeq 0.1$\cite{Th}. We are considering a  
larger value of $r$ in 
order to emphasize the presence of a gap in this figure.
\end{thebibliography}
\end{document}